\documentclass[prl,aps,showkeys,showpacs,superscriptaddress,twocolumn,%
amsmath,amssymb]{revtex4} 

\usepackage{graphics}
\usepackage{epsfig}
\usepackage{epstopdf}
\usepackage{amsmath}
\usepackage{setspace}
\usepackage[bookmarks=false,linkcolor=blue,urlcolor=black,colorlinks,citecolor=blue]{hyperref}
\usepackage{amsmath}
\usepackage{physics}

\begin{document}
\title{Stress Fluctuations in Transient Active Networks}
\author{Daniel Goldstein} 
\affiliation{Martin Fisher School of Physics, Brandeis University, Waltham, MA 02454, USA}
\author{Sriram Ramaswamy}
\affiliation{Centre for Condensed Matter Theory, Department of Physics, Indian Institute of Science, Bangalore 560 012, India}
\author{Bulbul Chakraborty}
\affiliation{Martin Fisher School of Physics, Brandeis University, Waltham, MA 02454, USA}

\date{\today}

\begin{abstract}
Inspired by experiments on dynamic extensile gels of biofilaments and motors, we propose a
model of a network of linear springs with a kinetics consisting of growth at a prescribed rate,
death after a lifetime drawn from a distribution, and birth at a randomly chosen node. The model
{captures} features such as the build-up of self-stress, that are not easily incorporated into
hydrodynamic theories. We study the model numerically and show that our observations can
largely be understood through a stochastic effective-medium model. The resulting dynamically
extending force-dipole network displays many features of yielded plastic solids, and offers a way to incorporate strongly non-affine effects into theories of active solids. A rather distinctive form for the stress distribution, and a Herschel-Bulkley dependence of stress on activity, are our major predictions.
\end{abstract}

\maketitle

\section{Introduction}

Conventional condensed matter  can be driven out of equilibrium by stresses and strains imposed at the boundaries.  
For example, most solids will flow if the external stress exceeds a yield stress, and suspensions can undergo shear-thinning or shear thickening.   Crosslinks in physical gels can be created and destroyed by external deformations leading to viscoelastic properties distinct from polymer melts~\cite{Tanaka_Edwards}.  
Active matter refers to condensed systems whose constituents are self-driven, that is, they are endowed with a local supply of free energy which they are equipped to convert to directed movement.
The question we address in this work is how  the  strains and stresses generated by active elements lead to transient, elastic networks.  Our primary interest is in the stress distributions and intrinsic rheological properties of such a network.  
This theoretical question has been motivated by experimental observations of spontaneous flow in an isotropic active gel, and the transition from turbulent to coherent flow when these systems are confined~\cite{Wu2017}.   

{
Understanding the development of large-scale flows in active matter is important for many biological systems.  The framework of active hydrodynamics\cite{review} { incorporates the effects of internally generated stress, and has been extraordinarily successful in describing the behavior of self-driven fluids~\cite{review}.  }
Active solids have been investigated, in the context of cytoskeleton reorganization~\cite{shiladitya_marchetti,debsankar_madan}, using hydrodynamic theories.  These theories predict the emergence of spontaneous oscillations and travelling waves when motor-driven stresses within the cytoskeleton exceed a threshold. 
%
Recent work~\cite{debsankar_madan} on the cytoskeleton as an active elastomer includes network rearrangements that reconfigure the connectivity of the actin filaments via crosslinks. 
This transient network approach is the active analog of theories of viscoelastic response in passive, unentangled but cross-linked polymer networks that reconfigure under
external deformations~\cite{Tanaka_Edwards}. {For a recent review of the rheology of active fluids see \cite{Saintillan_Annu_Rev}.}

%
Our work is motivated by a widely studied experimental system consisting of  microtubule bundles and motors\cite{Wu2017,BANS_dogic,Dogic_1,Dogic_2}.   The main features that are of interest to our theoretical work are: (i)  The active units are bundles of microtubules (MT) drawn together by depletion forces, and crosslinked by motor clusters that walk along them.  The motors induce relative sliding of microtubules with opposite polarity leading, in the experiments~\cite{Fiodor_thesis},  to extensile stresses and flow~\cite{Shelley}.
(ii) When no ATP is supplied to the motors the system behaves as a passive, isotropic  gel\cite{Wu2017}. (iii)
In the presence of ATP, the bundles exhibit a complex dynamics as they extend, bend, buckle, disassemble and reassemble\cite{BANS_dogic}.  (iv)  There is a bath of polarity-sorted bundles that are not extending, which become incorporated into the network of dynamical bundles.   

{There are features of this experimental system that are difficult to incorporate into existing hydrodynamic theories.  For instance,  unlike extensile force dipoles,  
 the MT bundles are dynamically extending  in length as they exert stresses on their environment.   These bundles can also induce states of self-stress in the network due to internal activity.  These are states in which stress builds up between two material points while the points remain in force balance, and thus do not move in space~\cite{Kane_Lubensky}.  Such states are also not easily incorporated into theories of active fluids.  In addition, as discussed above, network rearrangements are not naturally represented in theories of active solids.  Motivated by these features, we investigate a microscopic model of transient networks driven by internal extensile activity. 
 This is not meant to be a complete theory since we neglect hydrodynamic interactions and treat the MT bundles as ``ghost'' filaments without excluded volume interactions.   Our objective is to construct a piece of the theory that describes the stress fluctuations originating from active network reorganizations.   As we show in this paper, the flows that develop in these transient networks are characterized by stress distributions that are strongly non-Gaussian and whose temporal fluctuations are large and intermittent.  Any hydrodynamic theory of these flowing states should account for these {rather distinctive} stress fluctuations.}

\section{Model}  { We model the \textit{passive} system by a mechanically rigid \textit{elastic} network of springs~\cite{Kane_Lubensky}.}}
%
To model extensile activity, the elements of the network are represented as linear springs with spring constants $k$, whose equilibrium length, $s(t)$ {\it increases}  with time $t$.  The strength of the activity is measured by a single parameter $\alpha=\frac{d s(t)}{dt}$, which is the same for all springs.  These active springs represent the {dynamically extending MT bundles}  in the experimental system\cite{BANS_dogic,Shelley,Fiodor_thesis}.   
The observed disassembly and reassembly of the microtubule bundles suggest that a minimal model should be based on extensile dipoles that are ephemeral: active units can disappear from the network and new ones can be incorporated.  

{Measurements of the mechanical properties of a bundle with two filaments\cite{Fiodor_thesis} using optical traps show that antipolar arrangements can lead to both extensile and contractile forces.  However, at large motor concentrations the length of the bundle grows linearly with time.  This is the observation that motivates the linear growth of the equilibrium length of springs in our model.  When confined within a optical trap, the extending bundle buckles and ultimately fails.  Force measurements show that the force decreases with buckling before going to zero as the bundle breaks.  In our model, we do not include the softening of the spring preceding breakage but incorporate the nonlinearity only through the breakage after a prescribed lifetime, as detailed below.  Bundles with polar arrangements are observed not to  extend due to motor activity.  We envision these as existing as a bath of springs at their unextended equilibrium length that can become incorporated in the extensile network.}

If the active linear springs had infinite lifetimes, the  initial rigid network  would have the ability to support stresses without motion of the nodes.  As the equilibrium lengths of the springs increased, the stress in the network  would increase as the ``active'' strain accumulated, and the network would respond elastically\cite{Kane_Lubensky}.    In contrast, if the springs have a finite lifetime, the failure of springs will lead to network rearrangements.   
In the experimental system, the bundle disassembly process follows a complex dynamics\cite{BANS_dogic}.  In our minimal model, we simply assign a lifetime to each spring.
%

During its lifetime each spring exerts a force on the two nodes to which it is connected, proportional to the difference between its present length $\ell$ and its current equilibrium length $s(t)$.  The length of a spring is simply the distance between the two nodes it connects.
The nodes evolve with no inertia, i.e. the net force on a node generates a proportionate velocity, not an acceleration.

Network rearrangements are triggered by the ``death'' of a spring as it reaches its lifetime.  
At this point, the spring  is removed from the system and  a new spring with unit equilibrium length and unit extension is ``born'' and attached to a node picked at random from the network at a random angle.  The number of active springs is thus conserved.  As each spring is born it is assigned a lifetime $\tau$ picked from a distribution $P(\tau)$, to be specified below.
%

The ``death-birth''  mechanism alone would drive down the connectivity in the network since the new spring is attached to only one node. In order to enable the generation of a steady state with an average connectivity, we introduce a node-merging process.  Nodes that are within a {pre-specified merging radius $r_*$}, which is much smaller than the average separation of nodes, are merged and become a new node that inherits all the connections of its parent nodes. This mechanism by itself increases the connectivity of the system. The competition of the node merging and the death-birth of springs allows the network to reach a steady state with a well-defined distribution of node connectivity, as shown in the Supplementary Information (SI).   The evolution of the network to a steady state is illustrated in Fig. \ref{fig:NESS}. 

A simple, illuminating  example  of  how activity can lead to the development of self-stress and network rearrangements   is provided by a 1D network with all active springs having the same initial equilibrium lengths $s_0$ but a distribution of lifetimes. Since the activity $\alpha$ is constant across all springs, they will share the same equilibrium length $s_i(t)$ at time $t$. Each node will be in force balance as the stress increases: $\sigma \propto 2 k \alpha t$, where $k$ is the spring constant.  {\c Note that {$\sigma$  is the active stress, proportional to the growth-rate parameter $\alpha$, which encodes the strength of activity in our model.}  When the spring with the shortest lifetime fails it is replaced by an unextended spring at unit length at the same position due to the constraints of the one dimensional system. The spring forces on the nodes no longer balance, generating a flow of the nodes towards the ``youngest'' spring until the next spring breaks. The internal activity can thus lead {to ``yielding'' of the elastic solid in the absence of external stresses.  {Yielded disordered networks are known to  exhibit large-scale spatio-temporal heterogeneities of the stress.  As we show below,  the yielded active-spring network exhibits a  broad distribution of stresses in the steady state.} }
\begin{figure}[t]
			\includegraphics[width=\linewidth]{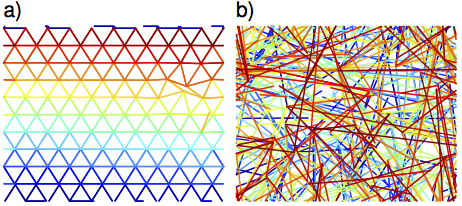}
\caption{\label{fig:model} { Snapshots illustrating  the transition from the initial network to the flowing, steady  state. (a) Snapshot just after the first spring has broken. The initial conditions of springs arranged in a triangular lattice with periodic boundary conditions is still seen through the majority of the system. Each spring is assigned a unique color to aid visualization.  (b) {A snapshot} in the steady state, which shows no memory of the initial triangular network. In the SI, we show the connectivity distribution of nodes in the steady state}}
\label{fig:NESS}
\end{figure}

%
%
%
%
%
%
%
%
\section{ Simulation of 2D network}
We evolve the system of active springs using  the following rules, which incorporate merging and breaking:
%
%
\begin{itemize}
\item [1]  Initial state:  Create  a triangular network of $N$ springs with unit equilibrium length, unit extension and lifetimes chosen from $P(\tau)$.
\item [2] Calculate the force on node $i$, ${\bf F}_i= \sum_{<ij>}  {\bf F}_{ij} = k \sum_{<ij>}(s_j-|\ell_j|)\hat{\ell_j}$ where the sum over $<ij>$ counts all the springs $j$ attached to node $i$.
\item [3] Move all nodes ${\bf r}_i(t+\Delta t)={\bf r}_i(t)+\mu {\bf F}_i  \Delta t $, where $\mu$ is the mobility.
\item  [4] If two nodes{, say $i$ and $k$,} are closer than the threshold merging distance, {$|{\bf r}_i-{\bf r}_k|<r_*$}, then merge the nodes. Create a new node at the average position of the two nodes that inherits all the connected springs, then delete the two old nodes.
\item [5] If a spring reaches its lifetime then remove it from the network and introduce a new spring at unit equilibrium length and no extension with one end point chosen randomly from all the nodes in the network and at a random orientation. {The lifetime of this new spring is also chosen from $P(\tau)$.}
\item [6] Delete any node that has no springs attached to it.
\end{itemize}
{The { fifth} rule erases any correlation between the lifetime of a spring and its spatial location in the network giving the model a mean-field flavor.   The internal stresses in this model, { which have elastic and active contributions},   are not spatially heterogeneous but display large temporal fluctuations.  If, however, the system is confined, then boundary conditions can lead to stress alignment and stress heterogeneities on length scales comparable to the average length that a spring grows to during its lifetime.}

We impose periodic boundary conditions in both directions. {  Unless otherwise stated, results are quoted for a rectangular box  with linear dimensions  $h= \frac{\sqrt{3}}{2} 50$, and $L=50$, which accommodates $N= 7500$ springs.  Our {network-rearranging} dynamics conserves the number of springs but not the number of nodes.}
{The distribution of lifetimes is constructed from a Gaussian distribution of  the maximum equilibrium length $s_{max}$ with mean $\expval{s_{max}}$ and a standard deviation comparable to the mean, which we choose to be $\expval{s_{max}}/4$.  Since the activity is the same for all springs, the lifetime is $\tau=(s_{max}-1)/\alpha$, therefore, $P(\tau)$ is also a Gaussian with mean $\langle \tau \rangle = \frac{\langle s_{max} \rangle -1}{\alpha}$ and standard deviation $(\alpha \langle \tau \rangle+1)/4\alpha$.  Tuning the activity, $\alpha$, thus changes the distribution of lifetimes and the network rearrangement dynamics speeds up with increasing activity, as observed in experiments~\cite{BANS_dogic}.  For any network realization we define the macroscopic stress tensor through the internal virial:  $\overleftrightarrow{\sigma} =\frac{1}{A} \sum_{\langle ij \rangle}  {\bf F}_{ij} \otimes ({\bf r}_i - {\bf r}_j)$.
To examine the viscous response of the system we apply a simple shear through the use of Lees-Edwards boundary conditions~\cite{LeeEd}.  }

 

\section{Numerical Results} 

We simulate the  model over  a range  of activities:  $\alpha=0.001$ to $\alpha=100$, {and study the properties of the steady-state that develops after an initial transient. {This steady state is usually reached after 2-3 $\langle \tau \rangle$, for $N$ springs with an average lifetime of $\langle \tau \rangle$, which means that on average, every spring has failed at least once before the system reaches steady state. }
We focus on the stress distribution in steady state. As is well established through theories of active hydrodynamics, these active stresses strongly influence flows.  Our approach is distinct from the studies based on hydrodynamics in that we (a) start from the elastic limit, (b) allow the extensile dipoles to grow in length, and (c) {incorporate a population of extensile dipoles within a fluctuating network.} These features lead to additional mechanisms of stress generation and instabilities.}

{We focus on the dynamical properties of the macroscopic stress tensor, since our network reorganization rules wash out any spatial heterogeneities.}  After the system is initialized, $\overleftrightarrow{\sigma}$ builds up from zero to a steady state value $\overleftrightarrow{\sigma}_{ss}$ about which it fluctuates. This behavior is illustrated  in Fig.\ref{fig:StressFigs}a. 
We can understand the approach to the steady state by considering how the initial triangular network responds to the activity.  As the equilibrium lengths of the springs and the stresses borne by them  increase, the nodes remain in force balance until the spring with the smallest lifetime breaks (Fig. \ref{fig:model}a). The network then rearranges with a sudden release of stress, as a new unstressed spring is introduced. 
At times long compared to the average lifetime of springs $\expval{\tau}$, the network reaches a steady state characterized by a time independent distribution of the ages of the springs (Fig. \ref{fig:model}b). The steady-state fluctuations are characterized by an approximately  linear rise in the stress between  death events, followed by nearly instantaneous stress drops. The average time between these stress drops  can be deduced simply by taking the ratio of the average lifetime  of a spring, $\langle{\tau}\rangle$,  to the number of springs in the system.

The stress evolution in the active spring model, summarized in Fig. \ref{fig:StressFigs},  resembles that observed in the yielding of elastic solids\cite{barrat_paper}. By thinking of the activity as a source of internal strain we can map time to accumulated strain in an elastic solid. In this mapping,  the transient of the average stress resembles the elastic branch of a solid and the steady state resembles the yielded state.  
{ 
Instead of being driven to flow by an imposed strain, the elastic network yields due to the internal strains generated by the active springs. {This internal driving has no macroscopic anisotropy. Therefore, as} shown in Fig. \ref{fig:StressFigs}(b),
{we measure} the trace of the  steady state stress tensor, ${\Pi}_{ss} = \rm{Tr} {\overleftrightarrow{\sigma}}_{ss}$, and find that it depends on activity $\alpha$ as $\Pi_{ss}=\Pi_{p}+\alpha^{\beta}$ where $\Pi_p$ is the pressure in the passive limit $\alpha\rightarrow0$.  We find $\beta \approx 1/2$:  a behavior similar to that observed in yield-stress fluids with $\alpha$ playing the role of  strain rate, and $\Pi_p$, the yield stress\cite{barrat_paper}. The relevant time-scale comparisons that determine whether this ``effective strain rate'' is small or large  is the ratio of the average lifetime of a spring $\langle \tau \rangle= \langle s_{max} \rangle /\alpha$ to the response time of the spring network $1/\mu k$. For $\langle \tau \rangle << 1/\mu k$, the springs cannot relax between network rearrangement events, which leads to a  larger stored stress in the network.
Fig. \ref{fig:SDEcomp} shows the distribution of $\Pi$ obtained from the time series in steady state. 

The active springs do not have a yield stress in the traditional sense  since their failure is controlled by a preassigned {lifetime; however,} the network rearrangements lead to stress reorganization and a distribution of effective yield-stresses emerges from the dynamics. We can define this effective yield stress, $\Pi_c$,  as the pressure (trace of stress tensor) of a spring at the moment it fails.  The distribution of  yield stresses in steady state, along with the steady-state distribution of $\Pi$ is shown in Fig. \ref{fig:StressFigs}(c).  It is seen that the distributions of both $\Pi$ and $\Pi_c$ are (i) broad and (ii) asymmetric.
{The asymmetry is a consequence of the linear-spring forces exerted by the extensile objects in our model. A spring can apply a contractile force if its local environment conspires to stretch is past its equilibrium length. These contractile forces result in negative stresses. The asymmetry of the stress distributions shows that on average these springs fail while exerting an extensile force.}    { We could have constructed an alternative model where the failure criterion of a spring was determined by its extension, which would be closer to the theories of passive transient networks~\cite{Tanaka_Edwards}.  In the experiments, however, different mechanisms of failure of the MT bundles are {observed. We therefore decided} to specify the lifetime distribution.   We believe that the qualitative features of the model  are determined by the finite lifetimes of the springs and not by  specific failure mechanisms.} 

{In the Supplementary Information, we show that a ``stress {automaton}''  that has been used to simulate plastic flow in soft glassy solids~\cite{barrat_paper,Picard}{ qualitatively reproduces} the stress distribution and the scaling of stress with activity found in the active spring model if we use as input the observed, emergent, yield stress distribution $\Pi_c$ (Fig. \ref{fig:stress}).  In {the main body of this work} we present an alternative stochastic, effective medium model that predicts the stress distribution from the dynamics of the springs without {recourse} to the yield-stress distribution.}

In steady state, springs with widely different lifetimes, $\tau$,  sample the same network environment, or conversely a spring with a given lifetime samples widely different network environments.  This lack of correlation between spatial location and lifetime suggests that  
the stress (yield-stress) distribution can be obtained by convolving the stress distribution, {\it conditioned on the age (lifetime) of a spring}, with the distribution of $t_b=t-t_0$, the time elapsed since the birth  of a spring.  Since every spring in the system {has} the same equilibrium-length growth rate $\alpha$,  the conditional stress distribution can be obtained algebraically from the conditional distribution of the extensions $P(\ell,t_b)$ of the springs.  

We can accurately fit the numerically measured $P(\ell,t_b)$ (Fig. \ref{fig:SDEcomp} (a)) by a Gaussian with mean $\mu=a t_b+1$ and variance $D=b_1 t_b+b_2 t_b^2$ (Fig. \ref{fig:SDEcomp} (c)). Of note is that this mean grows slower {than one would expect} from a single active spring. Through a change of variables {$\sigma(\ell,t_b)=(\alpha t_b + 1 -\ell)\ell$} we then obtain the conditional stress distribution.  
The calculation of the distributions of $\Pi$ and $\Pi_c$ by convolving this with the lifetime and the age distribution of the springs - both inputs to the system-  is presented in detail in the SI. Fig. \ref{fig:StressFigs}(c) compares the numerically measured stress distribution with the forms predicted by the above analysis, and demonstrates that our picture of the steady state applies.   The {distribution $P(\ell,t_b)$} reflects the different network environments that a spring samples during its lifetime, and is a measure of the disorder in the network  that results from the network rearrangements triggered by activity.  This is   the non-trivial ``emergent'' property that then controls the stress and yield-stress distributions.  This simplicity results from the mean-field character of our model, and we will use it in the next section to provide an effective medium theory of the stress fluctuations. }

We have also studied the response  of the active springs to  external deformations  by applying a simple shear strain through the use of Lee-Edwards boundary conditions\cite{LeeEd}. {For purposes of comparison with the unsheared case, we once again study $\Pi={\sigma_{xx}+\sigma_{yy}}$. As a rough indicator of how flow disrupts structure we measure the compressional ``viscosity'' $\nu_\Pi=\frac{\sigma_{xx}+\sigma_{yy}}{\dot{\gamma}}$ associated with the filaments}. As seen in Fig \ref{fig:StressFigs}(d), $\nu_\Pi$  scales as $\frac{1}{\dot{\gamma}}$ at low shear rates: $\dot{\gamma}\langle{\tau}\rangle<<1$.  This {weakening under shear}, reminiscent of a yielded plastic solid\cite{visc}, is a consequence of {the relaxation of} internal stresses stored in the active network in the absence of externally imposed shear.  In {addition,} our model predicts that $\nu_\Pi$ in the limit of $\dot{\gamma} \rightarrow 0$  increases with increasing activity $\alpha$.  { Since $\langle \tau \rangle$ decreases with $\alpha$,  this behavior is a direct consequence of the build up of larger stresses in the network ( Fig. \ref{fig:StressFigs} b). Viscosity defined as $\lim_{\dot{\gamma} \rightarrow 0} \sigma(\alpha,\dot{\gamma})/\dot{\gamma}$, therefore, increases with $\alpha$.} 

\begin{figure}[t]

		\includegraphics[width=\linewidth]{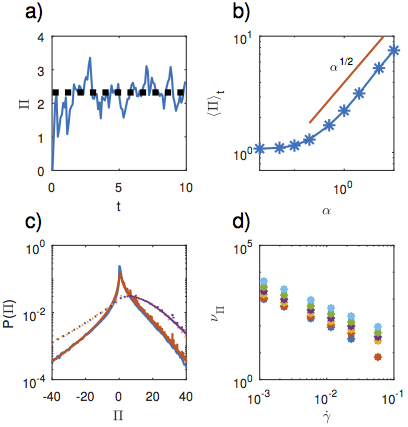}

\caption{\label{fig:StressFigs} (color online) (a) A typical time series of the pressure, $\Pi$,  illustrating  a linear stress build up and failure around the global `yield stress' of the system. (b) Steady state stress as a function of the activity. This behavior is well represented by Herschel-Bulkley law (see text) with activity taking  the place of the strain rate in a sheared amorphous solid~\cite{barrat_paper}.  (c) Distributions of $\Pi$  (blue solid) and $\Pi_c$ (purple dashed)  obtained from the simulations compared with those obtained from convolving the stress distributions with the age distribution (see text) (red solid, and orange dashed, respectively). (d) {``Viscosity''} as a function of imposed shear rate for different activities: $\alpha$ ranging from $0.1$ to $5$ bottom to top.
}
\label{fig:stress}
\end{figure}

{\section{Effective Medium Theory}}
Each spring in our active network is born with a lifetime picked from a quenched distribution.  In addition,  when a spring is born, it is attached to a node that is randomly picked from the network.  Thus, there is no correlation between the lifetime and the position of a spring.  In steady state, therefore, a spring with a given lifetime can be assumed to have sampled all possible network environments.  
%
  } {It is therefore natural to formulate an} effective medium theory for  the distribution $P(\ell,t_b)$. 
We begin by writing down the dynamics of the end points of a single spring that is not interacting with the network. As there are no externally imposed deformations, we can assume an isotropic system.  The dynamics of the {extension $\ell$ of a spring,} obtained using the force law and overdamped dynamics is: 
\begin{align}
\dv{\ell}{t_b}&=2 \mu k (\alpha t_b + s_0-\ell)
\end{align} 
{Solving this equation with $\ell(0)=s_0$ one finds $\ell(t_b)=s_0+\alpha t_b +\alpha(e^{-2 \mu k t_b}-1)/2\mu k$. There is a slow period for  $t_b << 1/2\mu k$ when $\ell(t_b) \simeq s_0 + O(t_b^2)$,  and the force exerted on the nodes $\approx  \mu k (\alpha t_b + O(t_b^2))$, which is $<< \alpha/2$.   For $t_b >> 1/2 \mu k$,  $\ell(t_b) \simeq s_0 +\alpha t_b -\alpha/{2\mu k}$,  which lags behind the equilibrium length.  Therefore, {the nodes attached to the spring feel} a force $\alpha/2$ away from one another.

In the active-spring network, the extension of a spring depends on its connectivity to the network, and the forces exerted by the other springs on the nodes that the spring is attached to.  Since a spring with a prescribed $t_b$ is born and samples the network completely randomly, given the rules of the model, we argue that the effect of the other springs  can be replaced by a ``noisy'' elastic medium.  {The extensile nature of the springs comprising this elastic medium can be incorporated by demanding that the elastic medium pushes out on its surroundings, on average.  This medium is thus characterized by an average force that resists compression,  and fluctuations of the elastic constant.   To motivate the mathematical model, we consider the extensional dynamics of a floppy spring ($\mu k=0$) embedded in an elastic medium that pushes on it:
\begin{align}
\pdv{\ell}{t_b}=-\ell \omega_{\eta}
\label{floppy_simple}
\end{align}
The extension will relax to zero as ${\ell}=\ell(0)\exp(-\omega_\eta t_b)$, with the characteristic timescale $1/\omega_{\eta}$ since the surrounding medium resists compression.  The noisy version of this system is represented by a Langevin equation with multiplicative noise:
\begin{align}
\pdv{\ell}{t_b}=-\ell \eta ~.
\end{align}
Here $\eta$ is a Gaussian white noise with $\expval{\eta}=\omega_\eta$, and a variance $\expval{\eta(t)\eta(t')}=2 \omega_c \delta(t-t')$ that characterizes the fluctuations of this elastic medium.
The effect of the multiplicative noise is to introduce a {\it random force}  that acts like a spring with a noisy spring constant and an equilibrium length of zero.
Making the substitution $\ell(t_b)=e^{z(t_b)}$ it can be seen that $z$ has the same distribution as a Brownian particle with a drift. As shown in the Supplemental Information (SI),  the solution for $z(t_b)$ can be used to calculate the extension:  $\expval{\ell}=\ell(0)\exp(-\omega_\eta t_b+\omega_c t_b)$.  On average, the effective medium compresses the floppy spring  with a characteristic time scale $\omega_\eta$, as in Eq. \ref{floppy_simple}. The noise in the effective medium, however, allows the two ends of the floppy spring to extend with a timescale $\omega_c$, mimicking the extensile activity of individual springs.   Thus in order to have a medium that resists being compressed on average  by  the extensile springs, we require $\omega_\eta>\omega_c$.  }

Generalizing to springs that are not floppy,   the dynamical  equation for $\ell$ is:}
\begin{align}
\pdv{\ell}{t_b}=2 \mu k (\alpha t_b +s_0- \ell)-\ell \eta ~.
\label{eq:langevin}
\end{align}
As seen clearly from Eq. \ref{eq:langevin}, the average compressive force exerted by the noisy medium competes with the intrinsic, extensile force due to the activity $\alpha$.
Details of the solution to Eq. \ref{eq:langevin}  is provided in the SI. Here, we present the results and compare them to our numerical simulations. The mean extension of a spring in this noisy elastic medium is found to be
\begin{widetext}
\begin{align}
\expval{\ell}=\frac{1}{(2 \mu k +\omega_\eta-\omega_c)^2}\qty[e^{-(2 \mu k +\omega_\eta-\omega_c)t_b}\qty[(\omega_\eta-\omega_c)(\omega_\eta-\omega_c+2 \mu k )s_0+2 \mu k  \alpha]+2 \mu k (2 \mu k +\omega_\eta-\omega_c)(s_0+t_b\alpha)]
\end{align}
\end{widetext}
For times $t_b(2\mu k +\omega_\eta-\omega_c)>>1$ the extension is:
\begin{align}
\expval{\ell}=\frac{2 \mu k \alpha}{2 \mu k+\omega_\eta -\omega_c}t_b~,
\label{long_time}
\end{align}
where the mean and variance of the noise now affect the speed at which these springs expand.   { Since $\omega_{\eta} \ge \omega_c$, the average extension is smaller than that of an isolated, extensile object ($\ell \approx \alpha t_b$). }

It is convenient to describe the system in terms of two characteristic {inverse} timescales, or rates:  (i) the pure spring elasticity scale, $\omega_{el}=2\mu k$, and $\omega_{eff}=(\omega_{el}+\omega_\eta -\omega_c)$, the scale characterizing the effective medium, which is a rearranging elastic network. Eq. \ref{long_time} in this representation gives:
\begin{align}
\expval{\ell}=\frac{\omega_{el}}{\omega_{eff}}\alpha t_b
\end{align}

The variance of $P(\ell,t_b)$ can similarly be calculated,  but the full form is less useful than the limits. We find that in the small $t_b$ limit the variance increases linearly with $t_b$ and in the large $t_b$ limit as {$t_b^2$:} 

\begin{align}
\lim_{t_b\rightarrow 0}\expval{\ell^2}-\expval{\ell}^2&\approx s_0^2 2 \omega_c t_b\label{eq:varlims}\\
\lim_{t_b\rightarrow \infty}\expval{\ell^2}-\expval{\ell}^2&\approx \frac{\omega_{el}^2 \omega_c}{(\omega_{eff} -\omega_c)\omega_{eff}^2} \alpha^2 t_b^2\label{eq:varlims2}
\end{align}

{As we can see from equations~\ref{eq:varlims} - \ref{eq:varlims2} and  Fig~\ref{fig:SDEcomp} (c), the effective medium theory predicts a distribution  whose long and short time limit of the mean and variance match the behavior of the measured conditional distributions from the simulation.  The distribution of $\Pi$, computed from the Langevin equations using values of $\omega_{\eta}$ and $\omega_c$ obtained from fitting $\langle \ell \rangle$  to simulations, compares well with the simulated distribution as shown in Fig. \ref{fig:SDEcomp} (b). The differences in the negative stress regime indicate that our model of the  effective medium as a purely extensile body, which exerts only compressive {forces,} does not accurately capture the tensile stresses in the springs.} 

{ In the SI, we present the numerically measured values of the noise parameters characterizing the  effective medium that represents the {steady state} of the active springs.  }
\begin{figure}[t]

	\includegraphics[width=\linewidth]{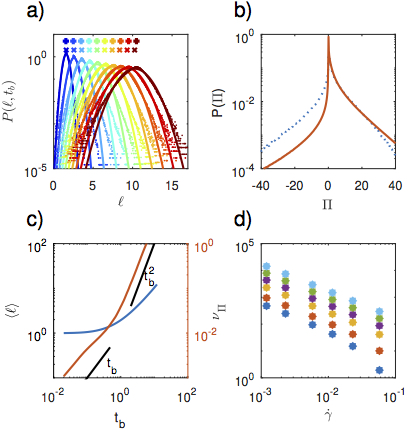}	
\caption{\label{fig:SDEcomp} (a)  - Spring-extension distributions for various $t_b$ with Gaussian fits. Blue corresponds to small $t_b$ and red to large $t_b$. the x's above the distribution denote the mean and the $+$'s represent the length of an isolated spring of the same age. Compare to the distribution obtained from the effective medium theory in Fig.~\ref{fig:EMTDdist} (b) Distributions of $\Pi$ obtained from simulations (blue-dashed) compared to the predictions of the effective medium theory (red-solid). (c) (blue) - The mean of the distribution of spring extension given that a spring has lived for exactly a time $t_b$.  (red)- The variance of the distribution of spring extension given that a spring has lived for exactly a time $t_b$. At small $t_b$ the variance grows linearly and crosses over to quadratic growth at large $t_b$  (d) Shear thinning behavior obtained from solving the stochastic differential equations ( Eq. \ref{eq:sde1} and \ref{eq:sde2}).
%
}
\end{figure}


To calculate the response to an externally imposed strain,  we modify the Langevin equations to  incorporate a simple shear in a two-dimensional representation. The shear rate $\dot{\gamma}$ is assumed to be small compared to the inverse of the average lifetime of a bundle: $\dot{\gamma} \langle \tau \rangle << 1$. 

\begin{align}\label{eq:sde1}
\dv{x}{t}=&\cos(\theta)\qty[2 \mu k (\alpha t_b +s_0- \ell)-\ell \eta]+\dot{\gamma} y\\
\dv{y}{t}=&\sin(\theta)\qty[2 \mu k (\alpha t_b +s_0- \ell)-\ell \eta]\label{eq:sde2}
\end{align}
{where} $\ell=\sqrt{x^2+y^2}$ and $\theta=\arctan(y/x)$. These equations are not analytically  tractable and must be solved numerically. We use the values of $\omega_\eta$ and $\omega_c$ obtained from simulations of the active spring model  at $\dot{\gamma}=0$. Using these parameters,  we find that $\nu_\Pi \propto \dot{\gamma}^{-1}$, as shown in  Fig \ref{fig:SDEcomp} (d), and the qualitative trend matches that of the simulations.

{ To summarize, replacing the  network of  active springs by an effective medium with ``noisy elasticity'' yields results for the spring dynamics that agree remarkably well with the full active spring simulations. This stochastic differential equation  reproduces the trends in viscosity and stress in the system and  the full stress distribution, qualitatively as shown in Fig. \ref{fig:SDEcomp}(b).  Furthermore the effective medium theory allows us to understand the nature of the stress fluctuations in the system as a result of the interaction of a spring with a noisy elastic medium that exerts compressive forces. }

\section{Conclusions}
Inspired by biological active networks we have created a model to explore the behavior of a network of dynamically extending  force dipoles. We have shown that this model shares several characteristics of a yielded plastic solid under periodic boundary conditions. We then show that an effective medium approach using a stochastic differential equation can reproduce key features of the behavior of the system including the viscosity and the behavior of the Herschel-Bulkley law.  {The mapping of the extensile transient networks to a noisy elastic medium offers an avenue for extending continuum theories of active solids~\cite{shiladitya_marchetti,debsankar_madan} to include strong nonaffine effects}.


\section*{Acknowledgments}
We acknowledge discussions with A. Baskaran, M. Hagan, Z. Dogic and D. Blair. DG and BC acknowledge discussions with K.  Ramola. . This research was supported in part by the Brandeis MRSEC: NSF-DMR 1420382, and by NSF PHY11-25915.  {SR acknowledges support from a J.C. Bose fellowship of the SERB, India, the Tata Education and Development Trust, the Simons Foundation and KITP.} BC's work has been supported by a Simons Fellowship in Theoretical Physics. We acknowledge the hospitality of the Kavli Institute for Theoretical Physics where part of this work was carried out.

\section*{Supplemental Information}
\label{sec:SI}
\begin{figure}[t]
	\includegraphics[width=.95\linewidth]{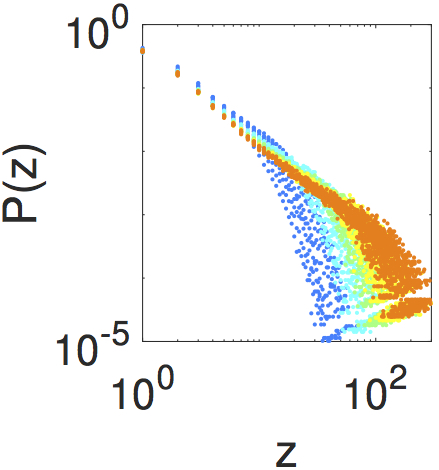}
	\caption{\label{fig:SIZ} P(z) for different  values of $\alpha$, and $s_{max}$. Colors correspond to different values of $s_{max}$: Blue - $s_{max}=10$, Teal - $s_{max}=20$, Yellow - $s_{max}=30$, Orange - $s_{max}=40$. The spread of points with a given color reflects the variation with activity for $\alpha=0.05\dots10$.    }
	\label{SI:fig1}
\end{figure}

\paragraph{Connectivity}
After a period of 2-4 spring lifetimes we find that the distribution of network connectivity, $P(z)$,   settles to a time-independent form, which can be represented by a power law with an  exponent of $~-1$ and an exponential cut off that depends on $s_{max}$. {A set of steady state connectivity distributions can be seen in Fig. \ref{fig:SIZ} for $s_{max}$ (denoted by color) ranging from $s_{max}=10\dots50$ and $\alpha=0.05\dots10$.  As seen in the figure, the cutoff in $P(z)$ is controlled primarily by $s_{max}$ with a much weaker dependence on the activity, $\alpha$.}

%

\paragraph{Stress distribution in steady state}
We find  that the probability of finding a spring of age $t_b$ at a length $\ell$ in the steady state $P(l,t_b)$, near the peak, can be fit by a Gaussian of the form:
\begin{align}
P(\ell,t_b)&=\frac{1}{\sqrt{ 2 \pi b t_b^{3/2}}}\exp(-\frac{(a t_b+1-\ell)^2}{ 2 (b_1 t_b+b_2t_c^{2}})
\end{align}
The equilibrium length of an active spring that has been growing for a time $t_b$ is $s=\alpha t_b+1$, which implies that the stress that has built up in the spring is: $\sigma(t_b,\ell)=k(\alpha t_b+1 - \ell)\ell$. Using this equation we can perform  a simple change of variables to obtain a conditional distribution for the stress.
\begin{widetext}
\begin{align}
P(\sigma,t_b)&=\frac{
	\exp \left(-\frac{\left(\sqrt{k} (2 a t_b-\alpha  t_b+2)+\sqrt{\alpha ^2 k t_b^2-4 \sigma }\right)^2}{8 k (b_1 t_b+b_2 t_b^{2})}\right)
	+\exp \left(-\frac{\left(\sqrt{k} (-2 a t_b+\alpha  t_b-2)+\sqrt{\alpha ^2 k t_b^2-4 \sigma }\right)^2}{8 k (b_1 t_b+b_2 t_b^{2})}\right)}
	{\sqrt{2 \pi } \sqrt{k(b_1 t_b+b_2 t_b^{2}) } \sqrt{\left| k t_b^2 \alpha ^2-4 \sigma \right| }}
\end{align}
\end{widetext}

\begin{figure}[t]
	\includegraphics[width=.95\linewidth]{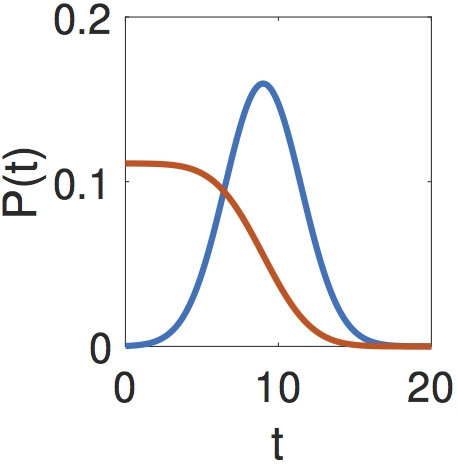}
	\caption{\label{fig:lifetimedist} Blue - Life time distribution, $P_l(t)$,  of springs {for $s_{max}=10$, $\alpha=1$, $\expval{\tau}=9$.  Red - The corresponding survival time distribution obtained from Eq. \ref{eq:survival}.   }}
\end{figure}

\begin{figure}[b]
	\includegraphics[width=.95\linewidth]{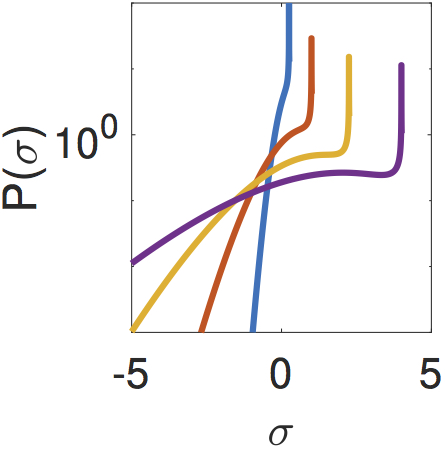}
	\caption{\label{fig:StressDist} $P(\sigma,t_b)$ for different values of $t_b$. Blue - $t_b=1$, Red - $t_b=2$, Yellow- $t_b=3$, Purple- $t_b=4$,}
\end{figure}

This stress distribution is doubly peaked. There is  a sharp peak at $\sigma=k t_b^2\alpha^2/4$ the maximum value the stress can have for a spring of a given age. The mean of this stress distribution is at $k\alpha t_b (a t_b - 1) -k(b_1 t_b+b_2 t_b^{2})$. Notably there is a large population under a lot of stress and the mean of this distribution increases with time. Negative stress here denotes that a spring is over extending and exerting a contractile force while positive stress denotes the spring is applying a extensile force. Some examples of these conditional distributions Are presented  in Fig. ~\ref{fig:StressDist}. By convolving this distribution with the survival probability, $P_s (t)$, of the springs in the system we can obtain the stress distribution in steady state. 

\begin{align}
P(\sigma)&=\int dt_b P_s (t_b) P(\sigma,t_b)~,
\label{eq:stress_disbn_ss}
\end{align}
The survival probability $P_s(t_b)$ and the  failure probability $P_f(t_b)$ are related to each other and to the {\it lifetime} distribution, $P_l (t)$, which is what we prescribe to define the model.  We deduce $P_s(t)$ from the prescribed Gaussian form of $P_l(t)$, as outlined below.  The probability that a spring will survive to an age $t_b+\delta t_b$ is equal to the probability that it has survived until age $t_b$ and then does not fail in a time interval $\delta t_b$.
\begin{align}
P_s(t_b+\delta t_b)=&P_s(t_b)(1-\delta t_b P_f(t_b))\\
\frac{1}{\delta t}(P_s(t_b+\delta t_b)-P_s(t_b))=&P_s(t_b)P_f(t_b)\\
\dv{P_s(t_b)}{t_b}=&-P_s(t_b)P_f(t_b)
\end{align}
By assigning a lifetime to each bundle,  we are prescribing the probability that a spring has failed exactly at time $t_b$, i.e., the probability that  a spring has survived until age $t_b$ then fails.  This distribution, $P_l(t_b) =P_s(t_b)P_f(t_b)$, we have prescribed to be a Gaussian:

\begin{align}
P_l(t_b) =\frac{e^{\frac{-(t_b-\tau)^2}{(2 (\alpha s_0+\tau)/4)^2)}}}{\sqrt{2\pi ((\alpha s_0+\tau)/4)^2}}~.
\end{align}
Using this form leads to a survival probability of
\begin{align}
P_s(t_b)=\frac{1-\erf(2^{3/2}\frac{t_b-\tau}{\tau})}{2\tau}
\label{eq:survival}
\end{align}

The stress distribution in steady state, which is independent of time, is obtained from Eq. \ref{eq:stress_disbn_ss} by using $P_s(t_b)$ obtained from the above equation, and the numerically evaluated distribution of extensions in steady state, $P(l,t_b)$, shown in  Fig. \ref{fig:SDEcomp}.   Additionally,  { by convolving the conditional stress distribution with the lifetime distribution, we can define an effective yield stress distribution}. This is not a true yield stress as putting more stress on a spring will not cause it to break, it is simply the stress springs are under at the moment of their death dictated by the lifetime assigned at birth. Fig. \ref{fig:StressFigs} compares the measured stress distribution with the forms predicted by the above analysis.  In the next section,  we present our solution to the stochastic differential equation representing the effective medium.

\paragraph{SDE Solution}
We provide details of the calculation of the stochastic differential equation that we have used to model $P(l,t_b)$. To get a sense of how a system behaves we first consider floppy limit of the spring, $k\rightarrow 0$  

\begin{align}
\dv{\ell}{t_b}=-\ell \eta
\end{align}
we can solve this equation though a change of variables $\ell=e^{y}$ now $\dv{\ell}{t_b}=\dv{y}{t_b}\ell$ and $\dv{y}{t_b}=-\eta$. Now if we want to find $\expval{\ell}=\expval{e^y}$ we can use a cumulant expansion to write down $log(\expval{\ell})=log{\expval{e^y}}=-a_\eta t_b+D t_b$. Thus $\expval{\ell}=e^{(-a_\eta+D)t_b}$, there is a relaxation timescale coming from the mean of the noise distribution as well as the variance. It is important to note here that this calculation was done using the notion of a Stratonovich~\cite{ito_stratonovich} integral so the notion of the chain rule remains the same in the presence of multiplicative noise. 

For a more complicated system of the form
\begin{widetext}
\begin{align}
\dv{\ell}{t_b}&=f(t_b)\ell+g(t_b) -\ell \eta(t_b)
\end{align}
One can make the transformation $\ell=z(t_b)\exp(\int_0^{t_b} f(s)ds+y(t_b))$ we can then write
\begin{align}
\dv{\ell(t_b)}{t_b}&=\dv{z(t_b)}{t_b}\exp(\int_0^{t_b} f(s) ds+y(t_b))+\ell(t_b)\qty[f(t_b)+\dv{y(t_b)}{t_b}]\\
\ell\dv{y(t_b)}{t_b}&+\dv{z(t_b)}{t_b}\exp(\int_0^{t_b} f(s) ds+y(t_b))=g(t_b)-\ell\eta(t_b)\\
&\mbox{Thus by making the identification}\\
\dv{z(t_b)}{t_b}&\exp(\int_0^{t_b} f(s) ds+y(t_b))=g(t_b)\\
z(t_b)&=\int_0^{t_b} g(s)\exp(-\int_0^{s} f(s') ds'-y(s)) ds
\end{align}
We can then arrive at
\begin{align}
\dv{y(t_b)}{t_b}&=-\eta(t_b)\\
\ell(t_b)&=\qty[\ell(0)+\int_0^{t_b} g(s)\exp(-\int_0^s f(s') ds'-\int_0^s\eta(s')ds') ds] \exp(\int_0^{t_b} f(s)ds-\int_0^{t_b} \eta(s)ds)\\
\ell(t_b)&=\qty[s_0+\int_0^{t_b} 2 \mu k (\alpha s+s_0)\exp(2 \mu k t_b+\int_0^s\eta(s')ds') ds] \exp(2 \mu k s -\int_0^{t_b} \eta(s)ds)
\end{align}
\end{widetext}

\begin{figure}[t]
	\includegraphics[width=.95\linewidth]{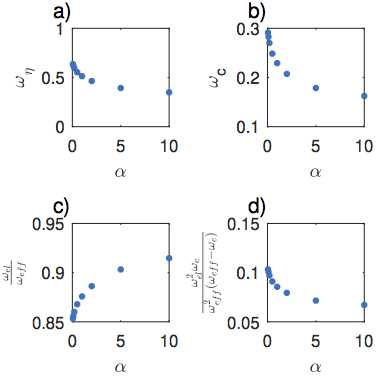}
	\caption{\label{fig:omegac} Values of noise parameters as a function of $\alpha$ for $s_{max}=50$. }
\end{figure}

\paragraph{Noise Parameters}
To complete the definition of our effective medium, we need to compute the parameters $\omega_{\eta}$ and $\omega_c$, which define the mean and variance of the noise.  We obtain the variation of these parameters with activity by fitting to results of simulations of the active spring model.   The easiest way to obtain the values of $\omega_c$ and $\omega_{\eta}$  is to use the long time limits of the mean and variance of $\ell$. Fitting the mean gives us a value for $\omega_{eff}$ and fitting the variance gives a value for $\omega_{c}$. It is then simple to solve for $\omega_\eta$. This process was carried out for values of $\alpha$ ranging from $\alpha=.01$ to $\alpha=10$ and $s_{max}=50$.   For smaller values of $s_{max}$, it was difficult to obtain adequate statistics since very few springs reached the long-time limit.   For each parameter value 20 different realizations were simulated and the values of $\omega_{\eta}$ and $\omega_{c}$ were averaged. A plot of these parameter for different values of $\alpha$ can be seen in Fig.~\ref{fig:omegac}.
 
\begin{figure}[t]
	\centering
	\includegraphics[width=.95\linewidth]{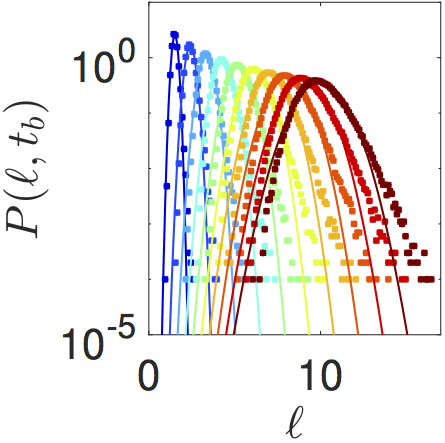}
	\caption[Conditional Length Distributions from the Effective Medium Theory ]{\label{fig:EMTDdist}A set of length distributions of the spring from the effective medium theory  conditioned by how long they have existed. Plots are shown for $t_b=(1,2,3,4,5,6,7,8,9,10)$ for a system where $s_{max}=10$. Gaussian fits of the regions near the peaks   are shown as solid lines.}
\end{figure}

A set of distributions of $\ell$ obtained from this effective medium theory can be seen in Fig.~\ref{fig:EMTDdist}.

\paragraph{Stress Automaton}
The stress fluctuations observed in our model are reminiscent of those observed in shear-driven amorphous solids, which undergo plastic failure.   A model that has been used to analyze plasticity in amorphous solids is a "stress automaton" in which the solid is divided into mesoscopic elements each characterized by a yield stress picked from a distribution.  Under externally imposed driving at a constant rate, the stress in each element, $\sigma_i$, builds up elastically until it reaches its yield stress.  Beyond this the stress decays to zero with a characteristic time $\tau$\cite{barrat_paper,Picard}.  The stresses in the other units are then altered by an amount $\delta \sigma_j = G_{ij} \sigma_i$ according to a prescribed Green's function $G_{ij}$.   The distribution of yield stresses and the Green's function parametrize the model.  

\begin{figure}[b]
	\centering
	\includegraphics[width=.7\linewidth]{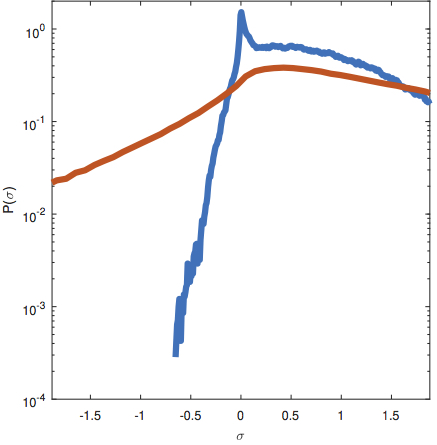}
	\caption{Distribution of $\sigma$ (blue line)  obtained from stress automaton simulations using the yield stress distribution computed from the active springs model with peak scaled to an arbitrary value (red line), which sets the stress scale for the automaton. These results were obtained for $\dot \gamma \tau = 1$.}
	\label{fig:automaton}
\end{figure}

We performed calculations using a  stress automaton under periodic boundary conditions using the elastic Green's function, and the yield stress distribution obtained from our numerical simulations of the active spring model (Fig. \ref{fig:stress} c).  Results were obtained for $\dot \gamma \tau = 1$ and $1/2$.
Comparing the results for $P(\sigma)$, shown in Fig. \ref{fig:automaton},  to $P(\Pi)$ shown in the main text, demonstrates  that the basic mechanism of failure and redistribution of stress  in an elastic medium captures the qualitative behavior observed in the active spring model including the cusp at small stress.    


%

\end{document}